\documentclass[11pt]{article}

\usepackage[parfill]{parskip}
\usepackage{fullpage}
\usepackage{bm}
\usepackage{graphicx}
\usepackage{natbib}
\usepackage{amsmath}
\usepackage{xcolor}
\usepackage{setspace}
\newcommand{\fw}[1]{\textcolor{black}{#1}} 
\newcommand{\track}[1]{{\color{black}#1}} 
\newcommand{\kw}[1]{\textcolor{black}{#1}} 

\title{Hybrid sample size calculations for cluster randomised trials using assurance}

\author{S. Faye Williamson$^{1}$\footnote{Corresponding author. Email: \href{mailto:faye.williamson@newcastle.ac.uk}{faye.williamson@newcastle.ac.uk}}, Svetlana V. Tishkovskaya$^{2}$, Kevin J. Wilson$^{3}$
\\ \small{$^{1}$ Biostatistics Research Group, Population Health Sciences Institute, Newcastle University, UK} 
\\ \small{$^{2}$ Faculty of Health and Care, Lancashire Clinical Trials Unit, University of Central Lancashire, UK}
\\ \small{$^{3}$ School of Mathematics, Statistics \& Physics, Newcastle University, UK}
}
\date{}
\usepackage{hyperref}                                   
\hypersetup{%
bookmarksnumbered,
pdfstartview={FitH},
linkcolor=blue,
citecolor=blue,
colorlinks=true,
pdfpagemode={UseNone},
plainpages=false
}%

\begin{document}


\maketitle

\begin{abstract} 

\noindent  \textbf{Background/Aims:} Sample size determination for cluster randomised trials (CRTs) is challenging because it requires robust estimation of the intra-cluster correlation coefficient (ICC). Typically, the sample size is chosen to provide a certain level of power to reject the null hypothesis in a two sample hypothesis test. This relies on the minimal clinically important difference (MCID) and estimates for the overall standard deviation, the ICC and, if cluster sizes are assumed to be unequal, the coefficient of variation of the cluster size. Varying any of these parameters can have a strong effect on the required sample size. \track{In particular, it is very sensitive to small differences in the ICC. A relevant ICC estimate is often not available, or the available estimate is imprecise due to being based on studies with low numbers of clusters.} If the ICC value used in the power calculation is far from the unknown true value, this could lead to trials which are substantially over- or under-powered. 

\noindent \textbf{Methods:} In this paper, we propose a hybrid approach using Bayesian assurance to determine the sample size for a CRT in combination with a frequentist analysis. Assurance is an alternative to traditional power, which incorporates the uncertainty on key parameters through a prior distribution. We suggest specifying prior distributions for the overall standard deviation, ICC and coefficient of variation of the cluster size, while still utilising the MCID. We illustrate the approach through the design of a CRT in post-stroke incontinence and compare the results to those obtained from a standard power calculation. 

\noindent  \textbf{Results:} We show that assurance can be used to calculate a sample size based on an elicited prior distribution for the ICC, whereas a power calculation discards all of the information in the prior except for a single point estimate. Results show that this approach \kw{can avoid misspecifying sample sizes when the prior medians for the ICC are very similar, but the underlying prior distributions exhibit quite different behaviour.} Incorporating uncertainty on all three of the nuisance parameters, rather than only on the ICC, does not notably increase the required sample size.

\noindent  \textbf{Conclusion:} Assurance provides a better understanding of the probability of success of a trial given a particular MCID and can be used instead of power to produce sample sizes which are more robust to parameter uncertainty. This is especially useful when there is difficulty obtaining reliable parameter estimates.

\end{abstract}

\section{Background} 

Cluster randomised trials (CRTs) are a type of randomised controlled trial (RCT) in which randomisation is at the cluster-level, rather than the individual-level as in standard RCTs. This means that \textit{groups} of individuals (e.g.\ general practices, schools or communities) are randomly allocated to different interventions (e.g.\ vaccination programmes or behavioural interventions). A common reason for implementing this design is to mitigate the risk of contamination {\track{or where individual randomisation is not feasible.}} 
Other justifications are detailed in \cite{eldridge2012practical}. 

Individuals within a cluster are likely to share similar characteristics (e.g.\ demographics), as well as be exposed to extraneous factors unique to the cluster (e.g.\ delivery of the intervention by the same healthcare professional). Consequently, outcomes from members of the same cluster are often correlated, which can be quantified by the intra-cluster correlation coefficient (ICC). This lack of independence reduces the statistical power compared to a standard RCT of the same size, meaning that the sample size needs to be inflated to allow for the clustering effect.

Various methods for sample size determination in CRTs exist (see \citealp{rutterford2015methods,gao2015sample}), which all rely on estimation of the ICC. In practice, ICC estimates are typically based on pilot studies, but these are often too small to provide precise and reliable estimates \citep{eldridge2016big}. An alternative simple approach is to use a conservative estimate of the ICC (e.g.\ the upper confidence interval limit) in the sample size calculation \citep{Browne}. However, this can lead to over-powered and unnecessarily large trials. A more reliable method is to combine ICC estimates from multiple sources, such as previous trials or databases listing ICC estimates \citep[e.g.][]{moerbeek2015power}, and use information on patterns in ICCs \citep[see][]{korevaar2021intra}. This raises further issues such as how to effectively combine the ICC estimates, how to adequately reflect their varying degrees of relevance to the planned trial and how to capture the uncertainty in the individual ICC estimates. \cite{lewis2021sample} suggest integrating over a range of possible ICC values, determined by confidence intervals obtained using methods in \cite{ukoumunne2002comparison}, to provide an `average’ sample size with respect to the ICC. However, this does not consider the uncertainty present in other design parameters, such as the treatment effect and variability of the outcome measures. Further, it assumes that each value of the ICC is equally likely.

Utilising a Bayesian approach for the trial design, in which prior distributions are assigned to the unknown design parameters such as the ICC, could further circumvent these issues and is particularly useful in settings where ICC estimates are not readily available. In the CRT literature, prior distributions for the ICC have been proposed based on subjective beliefs \citep{spiegelhalter2001bayesian} and single or multiple ICC estimates \citep{turner2004allowing}, which may be weighted by relevance of outcomes and patient population \citep{turner2005prior}. These are used to estimate a distribution for the power of the planned trial for a given sample size. Within the Bayesian framework, uncertainty in other design parameters can be incorporated into the sample size calculation in a similar way, and the relative likelihood of different parameter values is encompassed through specification of the prior distribution. For example, \cite{sarkodie2022hybrid} assign a prior to the overall standard deviation, in addition to the ICC, then describe a ‘hybrid’ approach to determine the sample size required to attain a desired ‘expected power’, defined as a weighted average of the probability that the null hypothesis is rejected (with weights determined by the priors). 

Hybrid approaches, which combine a Bayesian design with a frequentist analysis of the final trial data, have gained increasing popularity, particularly with respect to standard RCTs \citep{Kun21}. In this paper, we adopt a hybrid approach by using the Bayesian concept of \textit{assurance} to determine the sample size for a two-arm \track{parallel-group} CRT with a Wald test for the analysis. In contrast to traditional frequentist power, which represents a conditional probability that the trial is a success, given the values chosen for the design parameters and the hypothesised treatment effect, assurance typically refers to the \textit{unconditional} probability that the trial will be ‘successful’ \citep{chen2017statistical}. We modify this definition by conditioning on the minimal clinically important difference (MCID) instead of assigning a prior distribution to, and integrating over, the treatment effect \citep[as in][]{Kun21, ciarleglio2016selection}. This is more representative of the design stage of a trial, in which the treatment effect is typically fixed \textit{a priori} by investigators. Moreover, this ensures that the assurance will tend to one as the sample size increases so can be used analogously to traditional power, thus aiding interpretation. 

\fw{A key consideration} when applying a Bayesian design is how to specify suitable prior distributions. In contrast to the paper by \cite{sarkodie2022hybrid}, which assumes independent priors on the ICC and standard deviation, we suggest a joint prior distribution for these parameters, as described in Sections~\ref{PriorSpec} and \ref{FullAssurance_Prior}. In addition, we account for the fact that many CRTs have unequal cluster sizes by defining a prior distribution on the coefficient of variation of cluster size. This is often overlooked in standard sample size calculations for CRTs \citep{eldridge2006sample,zhan2021}. 

Our approach is motivated by a parallel-group CRT, Identifying Continence OptioNs after Stroke (ICONS), outlined in Section~\ref{ICONS_outline}. In Section~\ref{ICONS_redesign}, we illustrate the effects of redesigning this trial using the entire ICC prior distribution to inform sample size determination via an assurance calculation, rather than relying on a single point estimate from this distribution as in \cite{Tishkovskaya}. The impacts of varying the \kw{ICC prior distributions} on the chosen sample size are evaluated in Section~\ref{SensICC}. We perform sensitivity analyses on other design parameters in an additional simulation study \textcolor{black}{provided in the Appendix.}

\cite{jones2021bayesian} summarise the current state of play regarding the use of Bayesian methods in CRTs. In doing so, they highlight the “need for further Bayesian methodological development in the design and analysis of CRTs \ldots in order to increase the accessibility, availability and, ultimately, use of the approach.” This paper is therefore a timely contribution. 

\section{\fw{Methods}} 

\subsection{Analysis for \fw{CRTs}}

Suppose that we are designing a \fw{two-arm, parallel-group CRT assuming 1:1 randomisation of clusters and normally distributed outcomes}. \textcolor{black}{A common analysis} following the trial is to use a linear mixed-effects model. \fw{That is,} if $Y_{ij}$ is the response for individual $i=1,\ldots,n_j$ in cluster $j=1,\ldots,C$, then 
\begin{equation}\label{mixed}
Y_{ij} = \alpha + X_j\delta + c_j + e_{ij},
\end{equation}
where $\alpha$ is an intercept term; $X_j$ is a binary variable which takes the value 1 if cluster $j$ is allocated to the treatment arm and 0 if it is allocated to the control arm, \fw{so that $\delta$ represents the treatment effect;} $c_j\sim\textrm{N}(0,\sigma^2_b)$ is a random cluster effect \fw{with $\sigma^2_b$ denoting the between cluster variation} and $e_{ij}\sim\textrm{N}(0,\sigma^2_w)$ is the individual-level error \fw{with $\sigma^2_w$ denoting the within cluster variation}. This can be extended for stepped-wedge \fw{CRTs} by following the model in \cite{Hus07}.

The ratio of the variability between clusters $\sigma^2_b$ to \fw{the total variability $\sigma^2 = \sigma^2_b + \sigma^2_w$} determines the extent to which clustering \textcolor{black}{induces correlations between outcomes for individuals in the same cluster}. This is \fw{referred to as} the \textit{intra-cluster correlation \fw{coefficient}} (ICC), $\rho=\sigma^2_b/\sigma^2$ \citep{Kerry1998}.

The superiority of the treatment is assessed via a hypothesis test of $H_0:\delta\leq0$ versus $H_1:\delta>0$. Using a Wald test, assuming asymptotic normality, the test statistic is $Z=\hat{\delta}/\sqrt{\textrm{Var}(\hat{\delta})}$, where $\hat{\delta}$ is the estimate of $\delta$ \fw{and} $\textrm{Var}(\hat{\delta}) = 4\sigma^2[1+\{(\nu^2+1)\bar{n}-1\}\rho]/C\bar{n}$, \fw{where} $\bar{n}$ is the average sample size per cluster and $\nu$ is the coefficient of variation of cluster size\fw{, i.e.\ the ratio of the standard deviation of cluster sizes to the mean cluster size}.

\subsection{Choosing a sample size \fw{using assurance}}

The power of the \fw{one-sided} Wald test for significance level $\alpha$ can be approximated \citep{eldridge2016big} \fw{by}
\begin{equation}
P(n\mid\delta,\bm\psi) = \Phi\left(\delta\sqrt{\dfrac{C\bar{n}}{4\sigma^2[1+\{(\nu^2+1)\bar{n}-1\}\rho]}} - z_{1-\alpha}\right),
\label{PowerFunction}
\end{equation}
where $z_{1-\alpha}$ is the $100(1-\alpha)\%$ percentile of the standard normal distribution and $\bm\psi=(\sigma,\rho,\nu)$ is the vector of ``nuisance" parameters, excluding the treatment effect. For a two-sided Wald test, $z_{1-\alpha}$ \fw{would be} replaced by $z_{1-\alpha/2}$. \fw{For equal cluster sizes, the power function would take the same form as equation \eqref{PowerFunction}, with $\nu = 0$ and $\bar{n} = n_j = n $.}

In a standard power calculation, the sample size would be chosen as the smallest value which gives 80\% or 90\% power, based on values for $\bm\theta=(\delta,\bm\psi)$. The treatment effect $\delta$ could be specified as the MCID or an estimate based on a pilot study, similar historical trials or expert knowledge. The values used for $\bm\psi$ are typically estimates.

Alternatively, we can use assurance to choose the sample size. Whereas the power is conditioned on the chosen estimates for $\bm\psi$ and possibly $\delta$, the assurance represents the \textit{unconditional} probability that an RCT will achieve a successful outcome \citep{Oha01}. Assurance has been used \fw{almost} exclusively when the value to be used for $\delta$ is an estimate. In this case, suppose that the \fw{CRT} is a success if the null hypothesis is rejected \fw{by} the Wald test for $\delta$. Rather than using point estimates for $\bm\theta$, \fw{we could assign a} prior distribution $\pi(\bm\theta)$ \fw{to} it \fw{and define the assurance $A(n)$ as the power, averaged over \fw{the} uncertainty in $\bm\theta$}:
\begin{eqnarray}\nonumber
A(n) & = & \int_{\bm\theta}\Pr(H_0\textrm{ rejected}\mid\bm\theta)\pi(\bm\theta)d\bm\theta, \\
& = & \int_{\bm\theta}P(n\mid\bm\theta)\pi(\bm\theta)d\bm\theta. \label{eqn:ass}
\end{eqnarray}

One disadvantage of the assurance is that it tends to $\Pr(\delta>0)$ under $\pi(\delta)$ as the sample size increases. That is, unlike power, there may be no sample size for which the assurance is above the typical thresholds of 80\% or 90\%. \fw{\cite{Kun21} avoid this by conditioning the prior distribution for $\delta$ on $\delta>0$ in the assurance calculation. In this paper,} we consider the following alternative approach.

The assurance in (\ref{eqn:ass}) assumes that we choose $\delta$ in the sample size calculation based on {\em a priori} considerations of the likelihood of the treatment effect. \fw{Instead,} we consider the assurance in conjunction with a trial planned using the relevance argument, that is, using \fw{the} MCID for $\delta$, $\delta_M$. In this case, there is no need to define a prior distribution for $\delta$, and the assurance reduces to
\begin{eqnarray}\nonumber
A(n\mid\delta_M) & = & \int_{\bm\psi}P(n\mid\delta_M,\bm\psi)\pi(\bm\psi)d\bm\psi.
\end{eqnarray}
The advantage of this is that the assurance will now tend to one as the sample size increases.

To evaluate the assurance in practice, we sample values \fw{of} $(\bm\psi_j)_{j=1,\ldots,S}$ from the prior distribution $\pi(\bm\psi)$ for some large number of samples $S$, and use Monte Carlo simulation \fw{to} approximate the assurance as
\begin{eqnarray}
\tilde{A}(n\mid\delta_M) & = & \dfrac{1}{S}\sum_{j=1}^S P(n\mid\delta_M,\bm\psi_j), \nonumber \\ 
& \approx & \dfrac{1}{S}\sum_{j=1}^S \Phi\left(\delta_M \sqrt{\dfrac{C\bar{n}}{4\sigma_j^2[1+\{(\nu_j^2+1)\bar{n}-1\}\rho_j]}} - z_{1-\alpha}\right).
\label{MCsim}
\end{eqnarray} 

\subsection{Specification of priors}
\label{PriorSpec}

To evaluate the assurance\fw{,} we are required to specify a prior distribution for $\bm\psi$. This simplifies to specifying marginal prior distributions for each parameter if they can be assumed independent. Given that $\sigma^2$ and \fw{$\rho$ are both} functions of $\sigma^2_w$ and $\sigma^2_b$, it \fw{is} unlikely that $\sigma$ and $\rho$ can be assumed independent. \fw{Therefore,} we consider a joint prior distribution for $(\sigma,\rho)$ and a marginal prior distribution for $\nu$. In order for the assurance to be a meaningful representation of \fw{the} probability that the null hypothesis is rejected, these prior distributions should be informative, representing \fw{the} current state of knowledge about the possible \fw{parameter} values. This is an elicitation problem, and information to specify the priors can be obtained from relevant past data, expert knowledge or a combination (an example of this is given in Section~\ref{ICONS}).

\fw{Since} the coefficient of variation \fw{can only take} positive \fw{values}, a gamma distribution $\nu\sim\textrm{Gamma}(a_\nu,b_\nu)$ is a \fw{sensible choice} for its prior distribution. The \fw{hyperparameters} $a_\nu$ and $b_\nu$ could be chosen based on previous studies, via modelling or by eliciting expert knowledge \citep{eldridge2016big}.

One way to specify a joint prior distribution for $(\sigma,\rho)$ is \fw{to assign} independent priors to $\sigma^2_b$ and $\sigma^2_w$, \fw{which} will induce a correlation between $\rho$ and $\sigma^2$. 
If we sample values \fw{of} $\sigma^2_b$ and $\sigma^2_w$ from their priors, we can obtain samples from the joint prior of $(\sigma,\rho)$. Typical choices of prior distributions for $\sigma^2_b$ and $\sigma^2_w$ are (inverse) gamma distributions \textcolor{black}{because they provide conjugacy}. 

An alternative approach, relevant to our application, is to specify the joint distribution between $\rho$ and $\sigma$ directly. \fw{For example}, we can utilise a bivariate copula to encode the dependence between the parameters. A bivariate copula is a joint distribution function on $[0,1]^2$ with standard uniform marginal distributions \citep{Nel06}. It can be used to construct a joint prior for $\rho$ and $\sigma$ via
\begin{displaymath}
\pi_{\rho,\sigma}(\rho,\sigma) = \pi_\rho(\rho)\pi_\sigma(\sigma) c(u,v),
\end{displaymath}
where $\pi_\rho$ and $\pi_\sigma$ are marginal prior distributions, $c(u,v)$ is the bivariate copula density function \fw{evaluated at} $u=F_\rho(\rho)$ \fw{and} $v=F_\sigma(\sigma)$ for prior \fw{cumulative distribution functions (CDFs)} $F_\rho$ and $F_\sigma$. One simple choice is the Gaussian copula: 
\begin{displaymath}
c(u,v) = \dfrac{\partial^2}{\partial u\partial v} \Phi_\gamma(\Phi^{-1}(u),\Phi^{-1}(v)), 
\end{displaymath}
where $\Phi_\gamma$ is the CDF of the bivariate standard normal distribution with correlation $\gamma$, and $\Phi^{-1}$ is the inverse univariate standard normal CDF. The advantage of this structure is that it allows specification of the marginal prior distributions for $\rho$ and $\sigma$ separately to their dependence, which is given by $\gamma$.

\section{\fw{Application}} \label{ICONS}

\subsection{The ICONS post-stroke incontinence CRT} \label{ICONS_outline}

The approach developed in this paper \fw{is motivated by a planned parallel-group CRT, \track{``Identifying Continence OptioNs after Stroke"} (ICONS)}, which investigates the effectiveness of a systematic voiding programme \track{in secondary care} versus usual care on post-stroke urinary incontinence for people admitted to NHS stroke units \citep{thomas2015identifying}. The primary outcome is the severity of urinary incontinence \track{at three months post-randomisation}, measured using the International Consultation on Incontinence Questionnaire \citep{ICIQ21}. Although a feasibility trial, ICONS-I \citep{thomas2014identifying}, was conducted, \track{the resulting ICC estimate was of low precision and could not be used as a reliable single source to inform the planning of the proposed trial}.

ICONS therefore considered a Bayesian approach to combine multiple ICC estimates from 16 previous related trials. The opinions of eight experts regarding the relevance of the previous ICC estimates were elicited \citep{Hagan2019} and used to assign weights to each study and each outcome within a study. The elicited study and outcome weights were combined using mathematical aggregation \citep{Hagan2006} and incorporated into a Bayesian hierarchical model \fw{following the method in \cite{turner2005prior}}. The resulting constructed ICC distribution had a posterior median of $\hat{\rho} = 0.0296$ with a $95\%$ credible interval of $(0.00131, 0.330)$. Details of the expert elicitation process \track{and modelling} are described in \cite{Tishkovskaya}.

For the ICONS CRT, the sample size was chosen to give 80\% power with a 5\% significance level to detect $\delta_M = 2.52$ using a two-tailed independent-samples $t$-test and a common standard deviation $\sigma$ of 8.32 \fw{obtained from} the ICONS-I feasibility trial. The ICC was assumed to be less than or equal to $\hat{\rho} = 0.0296$. It was assessed as realistic to recruit between 40 and 50 stroke units, which required total sample sizes of $N = 480$ and $N = 450$, \fw{respectively}, and an \fw{average} sample size per cluster of $n=12$ and $n=9$, respectively. The \fw{original} sample size calculation assumed equally sized clusters. However, if we consider unequal \fw{cluster} sizes with $\nu=0.49$ (obtained from ICONS-I) and apply the Wald test, the required sample sizes remain the same.

\subsection{Redesigning the ICONS CRT using assurance} \label{ICONS_redesign}

We consider assurance as an alternative to power \fw{to determine the} sample size for the ICONS CRT. \fw{This} seems like a more natural approach given the uncertainty in the ICC and the extensive elicitation and modelling that was conducted to construct the ICC posterior distribution \textcolor{black}{(which forms the prior distribution for the assurance-based sample size calculation)}. Moreover, assurance incorporates the full ICC distribution into the sample size calculation, rather than relying on a single point estimate from it as in the power calculation.

\fw{We consider the following two forms of assurance.}

\subsubsection{Assurance based on the ICC prior \fw{only}}
\label{PartialAssurance_Prior}
\begin{figure}[ht]
\centering
\includegraphics[height=4in]{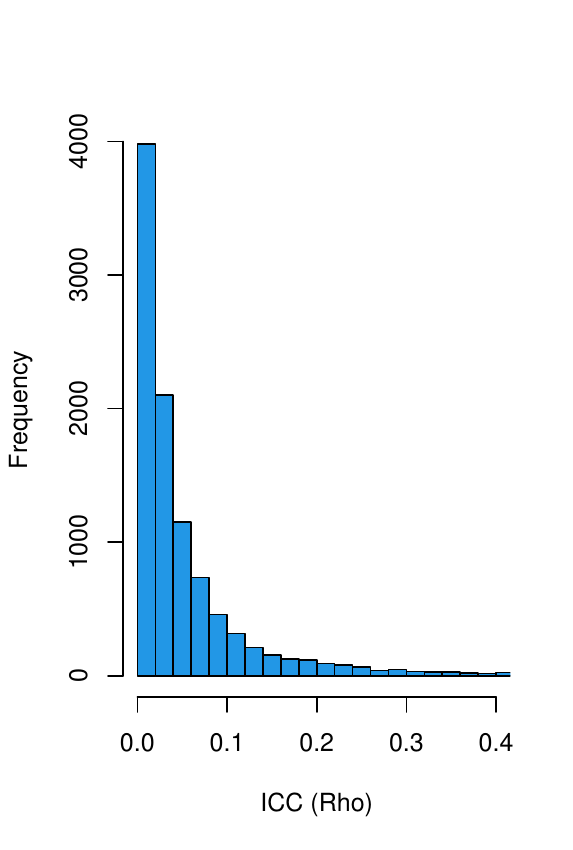}
\caption{Histogram of 10,000 samples of the ICC, $\rho$}
\label{hist}
\end{figure}

\fw{In the first case, we fix $(\sigma,\nu)$ using the point estimates obtained from ICONS-I and only consider the assurance with respect to the ICC.} We sample $S=10,000$ values of $\rho$ from its distribution (see Figure \ref{hist}) and approximate the assurance using (\ref{MCsim}). 
To obtain an assurance of 80\%, the resulting average sample sizes \fw{per cluster} are $\bar{n}=17$ for \fw{$C=40$ clusters ($C=20$ per arm) and $\bar{n}=11$ for $C=50$ clusters ($C=25$ per arm)}, requiring total sample sizes of $N=680$ and $N=550$, respectively (Table \ref{SUMMARY_sample_sizes}). Thus, the inclusion of uncertainty in the ICC results in a larger sample size than when using the posterior median ICC, but provides \track{a more realistic and robust study design}. Compared to the classical approach, the total sample size attained is smaller for the smaller number of clusters.

\begin{figure}[ht]
\centering
\includegraphics[width=0.45\linewidth]{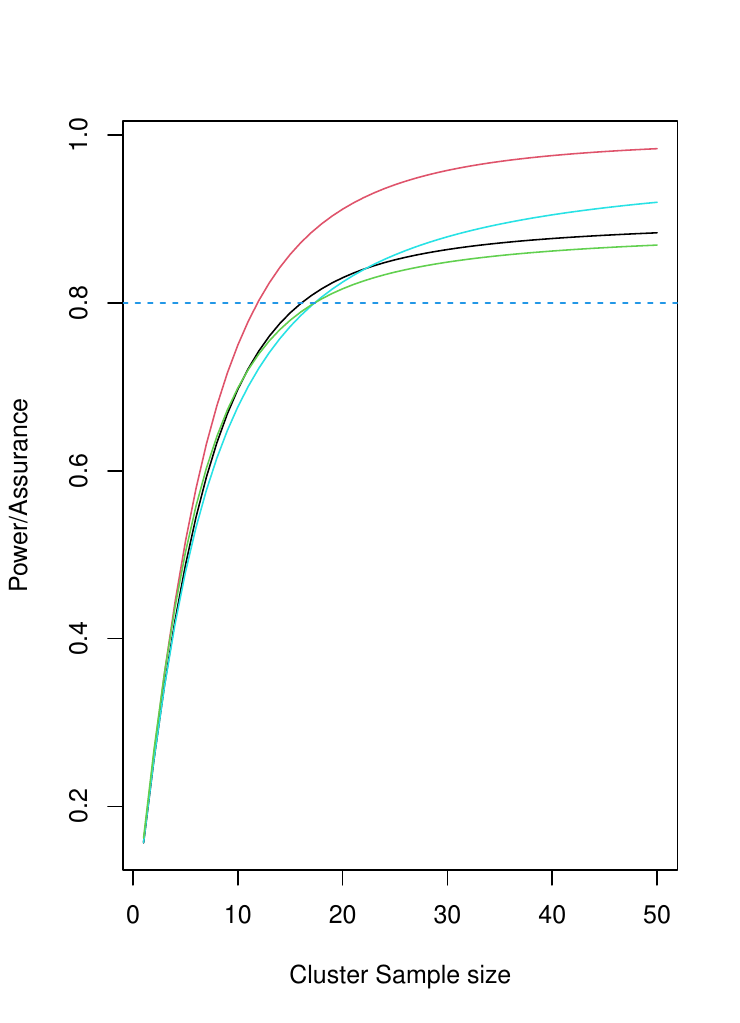}
\includegraphics[width=0.45\linewidth]{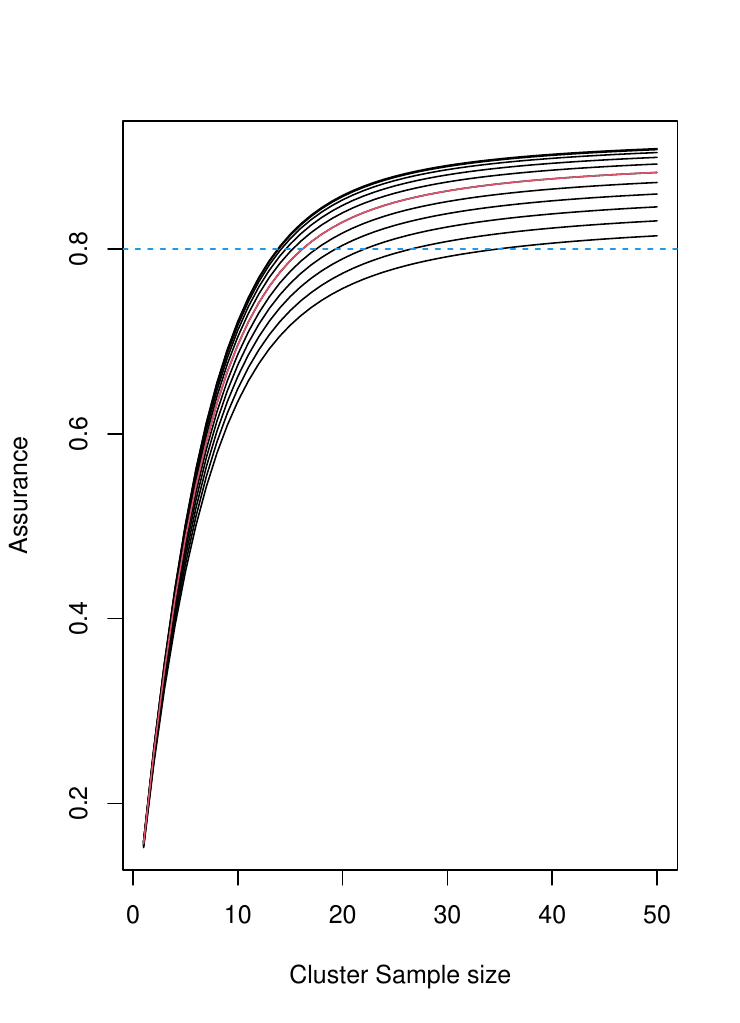}
\caption{Power and assurance curves for the \fw{ICONS CRT} (left). \textcolor{black}{The power using the posterior median ICC is red, the power using the median ICC from the 34 ICC estimates is light blue, the assurance with a prior only on the ICC is black and the assurance with a prior on all of the nuisance parameters $\bm\psi$ is green. The effect of varying the coefficient of variation $\nu$ on the assurance (right). $\nu$ varies between 0 (top curve) and 1 (bottom curve), with the red line at $\nu=0.5$. The horizontal line indicates the desired power/assurance. Each plot corresponds to} \fw{$C = 40$ ($C=20$ per arm)}.}
\label{trade}
\end{figure}

\fw{The left-hand side plot of Figure~\ref{trade} illustrates} the trade-off \fw{between cluster size and assurance/power}, for \fw{$C=40$ clusters ($C=20$ per arm)}. The power \fw{calculation based on the median from the elicited prior distribution of $\rho$} is represented by the red curve and the assurance \fw{with a prior on $\rho$ only} by the black curve. We see that the assurance requires a larger sample size \fw{than power when the target lies} above 0.5. \fw{We also include the power curve corresponding to the commonly used approach of taking the median of the 34 ICC estimates (blue line). For a target power of 0.8 (horizontal line), Figure~\ref{trade} shows that this method requires a larger sample size per cluster than the aforementioned methods.}

We \fw{illustrate} the effect of changing $\nu = (0, 0.1, \ldots, 1)$ on the assurance \fw{in the right-hand side plot of Figure \ref{trade}}. The red curve \fw{corresponds to} $\nu=0.5$, \fw{the top curve to $\nu = 0$ and the bottom curve to $\nu = 1$}. As $\nu$ increases, the assurance decreases for a given cluster size. We see that \textcolor{black}{the estimate of} $\nu$ has a relatively strong effect on the assurance, and hence the \fw{required} sample size. This implies that it needs to be estimated accurately, or its uncertainty should be taken into account in the assurance \fw{calculation}.

\subsubsection{Assurance based on the prior \fw{for $\psi$} }
\label{FullAssurance_Prior}

In the second case, we obtain the sample size required using \fw{an assurance calculation which averages} over a prior distribution on $\sigma$ and $\nu$, as well as the ICC. Using the data from \fw{ICONS-I}, we give $\sigma$ and $\nu$ gamma marginal prior distributions, centred at their estimated values of 8.32 and 0.49, respectively. The standard deviations of the prior distributions are chosen to represent a belief that $\sigma$ is very likely to be in the range $[5,11]$ and $\nu$ is very likely to be in the range $[0.3,0.7]$. Specifically, $\sigma\sim\textrm{Gamma}(a_\sigma,b_\sigma)$ and $\nu\sim\textrm{Gamma}(a_\nu,b_\nu)$, where $a_{\cdot}=m_{\cdot}^2/v_{\cdot}$ and $b_{\cdot}=m_{\cdot}/v_{\cdot}$, $m_\sigma=8.32,v_\sigma=1^2$ and $m_\nu=0.49,v_\nu=0.066^2$. 

To incorporate the dependence between $\rho$ and $\sigma$, we utilise the Gaussian copula with $\gamma=0.44$. This is chosen to be consistent with the correlation between $\rho$ and $\sigma$ that would result from independent prior distributions on the between and within group variances of $\sigma^2_b\sim\textrm{Gamma}(0.18,0.04)$ and $\sigma^2_w\sim\textrm{Gamma}(21.06,0.32)$, \fw{respectively}. The \fw{hyper}parameters of these two gamma prior distributions are chosen to provide the correct marginal means and variances for $\rho$ and $\sigma$. To sample values of $\rho$ and $\sigma$ from their joint prior distribution, we repeat the following steps:
\begin{enumerate}
    \item Sample $(x_i,y_i)$, $i=1,\ldots,S$ from $\textrm{N}_2(\bm 0,R)$, where $R$ is the prior correlation matrix with diagonal elements 1 and off-diagonal elements \fw{$\gamma=0.44$}.
    \item Calculate $(\rho_i,\sigma_i)$ as $\left(F^{-1}_\rho(\Phi(x)),F^{-1}_\sigma(\Phi(y))\right)$.
\end{enumerate}
The quantile function $F^{-1}_\sigma$ is that of the relevant normal distribution. The empirical quantile function $F^{-1}_\rho$ is used for $\rho$, based on the 10,000 prior samples.

\begin{figure}[ht]
\centering
\includegraphics[width=0.45\linewidth]{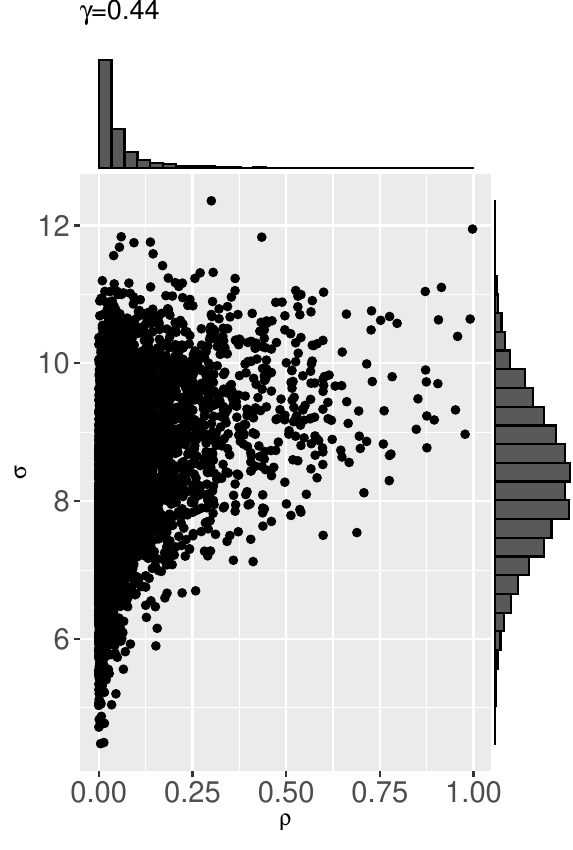}
\includegraphics[width=0.49\linewidth]{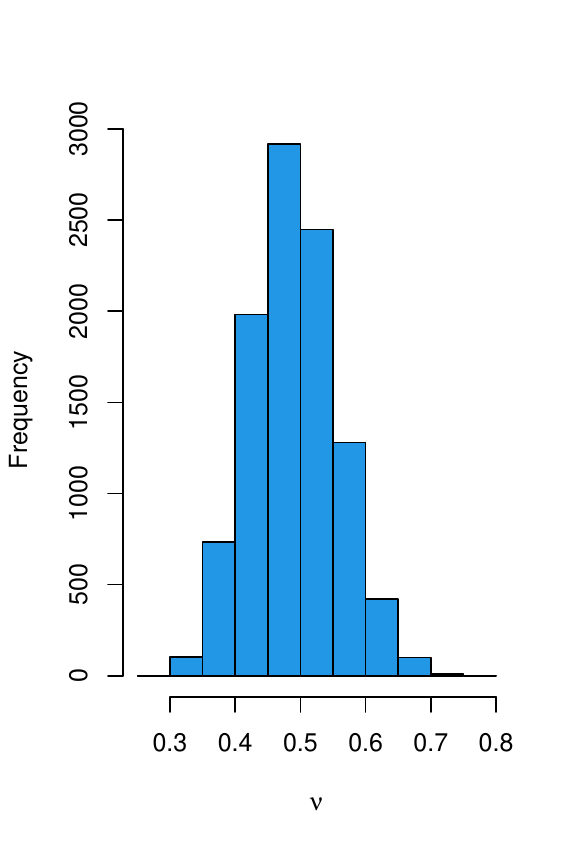}
\caption{The joint prior distribution between $\rho$ and $\sigma$ (left) and the marginal prior distribution for $\nu$ (right), based on 10,000 samples.}
\label{fig:priors}
\end{figure}

The resulting joint prior distribution for $(\sigma,\rho)$ and marginal prior distribution for $\nu$ are \fw{illustrated} in Figure~\ref{fig:priors}. We see that the marginal prior for $\rho$ remains as in Figure \ref{hist}, but the samples are positively correlated with the values of $\sigma$. 

The resulting average \fw{cluster} sample sizes for an assurance of 80\% are $\bar{n}=18$ for \fw{$C=40$ clusters ($C=20$ per arm)} and $\bar{n}=12$ for \fw{$C=50$ clusters ($C=25$ per arm)}, requiring total sample sizes of $N=720$ and $N=600$, respectively. By incorporating uncertainty on $\sigma$ and $\nu$, as well as $\rho$, \fw{the sample size increases only slightly, as illustrated in the left-hand side \fw{plot} of Figure~\ref{trade} (green line). To achieve a target assurance of 80\% (dashed horizontal line), the average sample size required per cluster increases from 17 to 18 when $C=40$; an increase in total sample size of approximately 5\%. }

Table~\ref{SUMMARY_sample_sizes} summarises the sample sizes required to attain a target power/assurance of \kw{80\%} for the various approaches \fw{applied to the ICONS trial}. ``Classical approach" refers to the multiple-estimate method of taking the median of the ICC estimates without taking the relevance of the different studies into account.
Relative to the classical approach that is often used in practice, the total sample size required when using the assurance-based method remains the same whilst incorporating uncertainty on all three parameters. 

\begin{table}[ht]
\centering
\begin{tabular}{|l|l|c|c|c|}
\hline
Method                                                                      & Priors                & \begin{tabular}[c]{@{}c@{}}Total number \\ of clusters, $C$\end{tabular} & \begin{tabular}[c]{@{}c@{}}Mean cluster \\ size, $\bar{n}$\end{tabular} & \begin{tabular}[c]{@{}c@{}}Total sample \\ size, $N$\end{tabular} \\ \hline
\begin{tabular}[c]{@{}l@{}}Power\\ (classical approach)\end{tabular}        & NA                    & \begin{tabular}[c]{@{}l@{}}50\\ 40\end{tabular}                          & \begin{tabular}[c]{@{}l@{}}10\\ 18\end{tabular}                         & \begin{tabular}[c]{@{}l@{}}500\\ 720\end{tabular}                 \\ \hline
\begin{tabular}[c]{@{}l@{}}Power\\ (based on posterior median)\end{tabular} & NA                    & \begin{tabular}[c]{@{}l@{}}50\\ 40\end{tabular}                          & \begin{tabular}[c]{@{}l@{}}9\\ 12\end{tabular}                          & \begin{tabular}[c]{@{}l@{}}450\\ 480\end{tabular}                 \\ \hline
Assurance                                                                   & $\rho$                & \begin{tabular}[c]{@{}l@{}}50\\ 40\end{tabular}                          & \begin{tabular}[c]{@{}l@{}}11\\ 17\end{tabular}                         & \begin{tabular}[c]{@{}l@{}}550\\ 680\end{tabular}                 \\ \hline
Assurance                                                                   & $\bm\psi = (\sigma, \rho, \nu)$ & \begin{tabular}[c]{@{}l@{}}50\\ 40\end{tabular}                          & \begin{tabular}[c]{@{}l@{}}12\\ 18\end{tabular}                         & \begin{tabular}[c]{@{}l@{}}600\\ 720\end{tabular}                 \\ \hline
\end{tabular}
\caption{Summary of sample sizes obtained for the ICONS CRT based on power and assurance calculations.}
\label{SUMMARY_sample_sizes}
\end{table}

\subsection{Sensitivity \fw{analysis for the ICC prior}} 
\label{SensICC}

In the above, we consider the ICC prior distribution based on all \fw{eight} reviewers and all 16 relevant studies. In this section, we investigate the sensitivity of the \fw{assurance-based sample size (with priors on $\bm\psi$)} to varying assumptions on the reviewers and relevant studies, and compare this to the sensitivity of the \kw{sample sizes from} power calculations \fw{(using the posterior median ICC)}.

To recognise uncertainty in the individual reviewers’ responses, and in how these responses were pooled, the mathematical aggregation was refitted with alternative reviewer importance weights:\ equal weights of 0.125 for all reviewers and \fw{using} a rank sum approach \citep[see][]{Tishkovskaya}. For the rank sum approach, we use Cronbach’s alpha score and assign ranks to each reviewer according to this score. In addition, we \fw{rerun} the Bayesian hierarchical model for only the top \fw{4 (25\%), 8 (50\%) and 12 (75\%)} most relevant studies. \fw{We refer to the five variations of the original ICC prior distribution as: equal weights, differentiated weights, top 4, top 8 and top 12}. 

\begin{figure}[ht]
    \centering
    \includegraphics[width=0.48\linewidth]{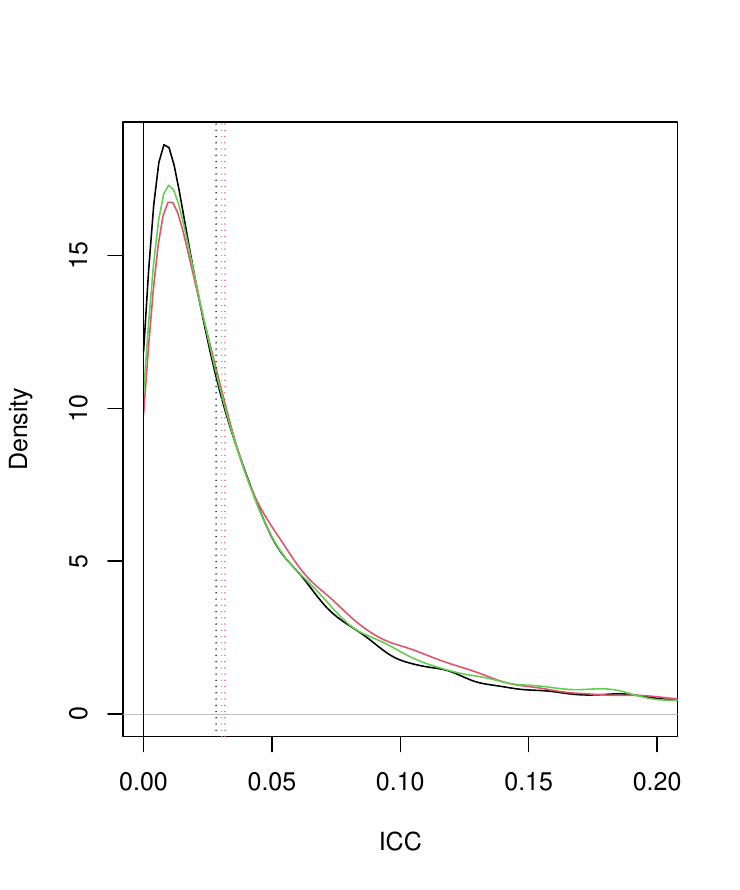}
     \includegraphics[width=0.48\linewidth]{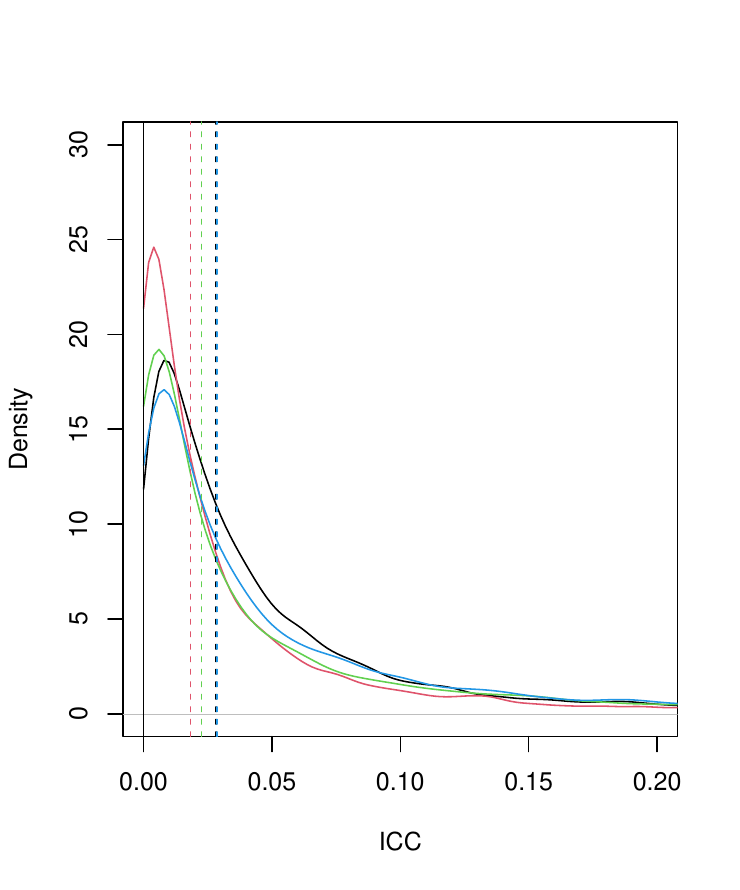}
    \caption{Left:\ The densities of the differentiated weights (red), equal weights (green) and original ICC prior (black). Right:\ The densities of the top 4 (red), top 8 (green), top 12 (blue) and original ICC prior (black). \kw{The prior medians are represented by vertical dashed lines.}}
    \label{fig:sens}
\end{figure}

The differentiated weights prior (red) and equal weights prior (green) are provided alongside the original prior \fw{(black)} in the left-hand side \fw{plot} of Figure \ref{fig:sens}. The top 4 prior (red), top 8 prior (green) and top 12 prior (blue) are given alongside the original prior \fw{(black)} in the right-hand side \fw{plot} of Figure \ref{fig:sens}. \kw{In both plots, the prior medians are given by vertical dashed lines.}

We see that the ICC prior remains similar to the original prior whether differentiated weights or equal weights are used, \kw{although both alternative weightings assign more probability to the ICC taking larger values}. \fw{There} is a larger change when using the top 4, top 8 or top 12 studies. In each case, the alternative prior is more diffuse than the original prior. \kw{Relatively large changes in the prior can cause only small changes in the prior median (e.g.\ the original prior compared to the top 12 prior).} The effects of the alternative ICC priors on the sample sizes are shown in Table~\ref{tab:sens}.

\begin{table}[ht]
\centering
\begin{tabular}{|l|cc|cc|cc|cc|}\hline 
& \multicolumn{4}{c|}{$C=50$} & \multicolumn{4}{c|}{$C=40$}  \\
\hline
& \multicolumn{2}{c|}{Assurance} &  \multicolumn{2}{c|}{Power} & \multicolumn{2}{c|}{Assurance} &  \multicolumn{2}{c|}{Power} \\
ICC Estimate/Prior & $\bar{n}$ & $N$ & $\bar{n}$ & $N$ & $\bar{n}$ & $N$ & $\bar{n}$ & $N$ \\ \hline
Original & 12 & 600 &  9 & 450 & 18 & 720 & 12 & 480   \\
Differentiated weights & 13 & 650 & 10 & 500 & 20 & 800 & 13 &520  \\
Equal weights & 13 & 650 &  9 & 450 & 19 & 760 & 13 & 520  \\
Top 4 & 12 & 600 & 8 & 400 & 18 & 720 & 11 & 440  \\
Top 8 & 14 & 700 & 9 & 450 & 23 & 920 & 12 & 480   \\
Top 12 & 15 & 750  & 9 & 450 & 24 & 960 & 12 & 480 \\ \hline
\end{tabular}
\caption{\textcolor{black}{The average sample size per cluster $\bar{n}$ and the total sample size $N$ required for the \fw{ICONS CRT} using assurance (with priors on $\bm\psi$) and power based on the original \fw{ICC} estimate/prior and five alternative estimates/priors when $C=50$ and $C=40$. The power is based on the posterior median of the ICC.}}
\label{tab:sens}
\end{table}

\kw{We see smaller changes in sample sizes for $C=50$ than $C=40$ using assurance. Overall, we observe larger changes in sample size using assurance than power based on the prior median of the ICC. This illustrates the risk with using just the median; it takes no account of the prior probability that the ICC could be relatively large, so has the potential to systematically underestimate the required sample size. In contrast, the assurance-based sample size is sensitive to the entire ICC prior distribution, particularly the upper tail.}

\kw{To illustrate this point, compare the original ICC prior (black) to the top 12 prior (blue) in the right hand side of Figure \ref{fig:sens}. They have substantially different priors, resulting in large differences in sample sizes required under assurance (600 versus 750 when $C=50$, respectively). However, their prior medians are almost identical, resulting in identical sample size requirements under power (450 when $C=50$).}

\textcolor{black}{In the Appendix, we further evaluate the properties of the hybrid approach compared to power via a simulation study.}

\section{Conclusions}

A standard sample size calculation requires pre-specification of parameters that are unknown at the design stage of a trial. Unique to sample size calculations for typical CRTs is the ICC, which requires robust estimation to avoid over- or under-powering the trial. Unnecessarily high ICC values, for example, lead to inefficient trials, increasing the number of clusters and/or participants, and overall trial costs. In practice, parameter uncertainty is typically not considered, which can be problematic given the sensitivity of the sample size to small differences in the ICC.

This paper proposes an alternative approach to sample size determination for CRTs using the Bayesian concept of assurance to incorporate parameter uncertainty into the design. The advantage of this approach is that it yields designs that provide adequate power across the likely range of parameter values and is therefore more robust to parameter misspecification. This is particularly important when there is difficulty obtaining a reliable ICC estimate, such as in the ICONS post-stroke incontinence CRT used to motivate this work.

We assign prior distributions to the ICC, overall standard deviation and coefficient of variation of the cluster size, whilst setting the treatment effect equal to the MCID in line with standard practice. We consider a joint prior for the ICC and standard deviation to model the dependency between these parameters. In the motivating case-study, we use the entire ICC prior distribution elicited from expert opinion and data from previous studies to inform the sample size. Further work could consider using a commensurate prior to synthesise multiple sources of pre-trial information on the ICC, as in \cite{Haiyan2022}.

Sensitivity analyses of the assurance-based sample size to different ICC priors showed that different behaviour of the prior, particularly in the upper tail, can have quite a strong effect on the required sample size. Using a point estimate from this prior, for example the median, can miss this overall behaviour and result in sample sizes which are systematically too small, based on current knowledge about the ICC. Additional sensitivity analyses conducted on the overall standard deviation showed that the greater the uncertainty expressed in the prior, the more robust the assurance-based sample size is (see Appendix).

Uncertainty in the treatment effect can also be incorporated into the assurance calculation in a similar way. This may be appropriate for non-inferiority trials, for example, where the non-inferiority margin is fixed in advance and the treatment difference can be considered a nuisance parameter.

In line with regulatory requirements, we have maintained a frequentist analysis to present a hybrid framework. Further work could consider a fully Bayesian approach by using assurance when the success criterion is based on the posterior distribution of the treatment effect \citep[e.g.][]{spiegelhalter2001bayesian}

The hybrid approach presented in this paper can be applied to avoid incorrectly powered studies resulting from ill-estimated model parameters, to mitigate the impact of uncertainty in the ICC and other nuisance parameters, and to incorporate expert opinion or historical data when designing a CRT.


\bibliographystyle{apalike}
\bibliography{refs}

\end{document}